\documentclass[size=10.5pt, paper=A4]{scrartcl}

\usepackage[margin=1in]{geometry}
\usepackage{placeins}
\usepackage{float}
\usepackage[doublespacing]{setspace}

\usepackage{graphicx}
\usepackage{transparent}
\usepackage{import}
\usepackage{xr-hyper}
\usepackage{hyperref}
\usepackage{xcolor}
\usepackage{booktabs}
\usepackage{enumitem}
\usepackage{placeins}
\usepackage{pdfpages}

\usepackage[backend=bibtex,
style=nature,
doi=false,
url=false, 
isbn=false,
date=year,
defernumbers=true,
]{biblatex}

\usepackage{caption}
\DeclareCaptionLabelSeparator{dot}{. }
\usepackage{amsmath}

\usepackage{etoolbox}
\usepackage{authblk}

\newcommand\CoAuthorMark{\footnotemark[\arabic{footnote}]}

\AtBeginEnvironment{picture}{\sffamily\fontsize{8}{9.6}\selectfont\mathversion{sansmath}}
\AtBeginEnvironment{pgfpicture}{\mathversion{sansmath}}
\AtBeginEnvironment{picture}{\fontsize{8}{9.6}\selectfont{}}

\begin{document}
	
\title{Strong Repulsive Lifshitz-van der Waals Forces on Suspended Graphene}

	\author[1]{Gianluca Vagli \footnote{These authors contributed equally to this work.}}
\author[1]{Tian Tian \protect\CoAuthorMark}
\author[1]{Franzisca Naef}
\author[1]{Hiroaki Jinno}
\author[1]{Kemal Celebi}
\author[2,3]{\par Elton J. G. Santos}
\author[1]{Chih-Jen Shih \thanks{Corresponding author. Email: chih-jen.shih@chem.ethz.ch}}
\affil[1]{Institute for Chemical and Bioengineering, ETH Z{\"{u}}rich,  CH-8093 Z{\"{u}}rich, Switzerland}
\affil[2]{Institute for Condensed Matter Physics and Complex Systems, School of Physics and Astronomy, The University of Edinburgh, EH9 3FD, UK.}
\affil[3]{Higgs Centre for Theoretical Physics, The University of Edinburgh,  EH9 3FD,  United Kingdom}
\date{} 	
	\maketitle{}
	
	\pagebreak{}
	
\begin{quote}
		\bfseries{
		Understanding surface forces of two-dimensional (2D) materials is of fundamental importance as they govern molecular dynamics and atomic deposition in nanoscale proximity. Despite recent observations in wetting transparency and remote epitaxy on substrate-supported graphene, very little is known about the many-body effects on their van der Waals (vdW) interactions, such as the role of surrounding vacuum in wettability of suspended 2D monolayers. Here we report on a stark repulsive Lifshitz-van der Waals (vdW) force generated at surfaces of suspended 2D materials, arising from quantum fluctuation coupled with the atomic thickness and birefringence of 2D monolayer. In combination with our theoretical framework taking into account the many-body Lifshitz formulism, we present direct measurement of Lifshitz-vdW repulsion on suspended graphene using atomic force microscopy. We report a repulsive force of up to 1.4 kN/m$^2$ at a separation of 8.8 nm between a gold-coated AFM tip and a sheet of suspended graphene, more than two orders of magnitude greater than the Casimir-Lifshitz repulsion demonstrated in fluids. Our findings suggest that suspended 2D materials are intrinsically repulsive surfaces with substantially lowered wettability. The amplified Lifshitz-vdW repulsion could offer technological opportunities such as molecular actuation and controlled atomic assembly.}
	\end{quote}
	
	\section*{Introduction}
	
	When two electroneutral objects are brought in proximity in a polarizable medium, the correlations in their temporal electromagnetic (EM) fluctuations usually lead to an attractive interaction\cite{Woods_2016_vdW_review}. 
	At small separations ($<$ 10 nm), this interaction is known as the vdW forces\cite{Parsegian_2010_vdW}, while at large separations ($>$ 20 nm) where the retardation effect comes into play, it is termed the Casimir forces\cite{Casimir_1948_first,Casimir_1948_second}.
	Early vdW theories\cite{Keesom_1915_vdW, Maitland_1981_intermolecular, London_1937_vdW} assumed the total interaction between two objects, each consisting of many molecules, is simply the sum of intermolecular potentials, which ignored the fact that the intermolecular interactions are strongly screened by the surroundings. 
	By considering macroscopic properties using quantum field theory and statistical physics, seminal work presented by Lifshitz et al. \cite{Dzyaloshinskii_1961_lifshitz} completely abandoned the pairwise additive assumption and predicted that quantum fluctuations can lead to repulsive interactions in both vdW and Casimir regimes. The existence of Casimir repulsion was later experimentally verified in several fluid-based systems \cite{Munday_2009_afm,Feiler_2008_superlubri,Zhao_2019_casimir_trap, Tabor_2011_rep_air, Schmidt2022}. 

	Consider two semi-infinite three-dimensional objects, \textit{A} and \textit{B}, interacting across a polarizable medium, \textit{m}.
	As the interaction potential in the Lifshitz
	theory\cite{Dzyaloshinskii_1961_lifshitz,Parsegian_2010_vdW} is
	proportional to the product of effective polarizabilities of \textit{A} and \textit{B}
	screened by \textit{m}, the most straightforward approach to generate Casimir or vdW
	repulsion is to design a set of materials such
	that\cite{Munday_2009_afm, GongCorradoMahbubSheldenMunday+2021+523+536}
	\begin{equation}
		\label{eq:eps-ineq}
		(\varepsilon_{\mathrm{A}} - \varepsilon_{\mathrm{m}})(\varepsilon_{\mathrm{B}} - \varepsilon_{\mathrm{m}}) < 0
	\end{equation}
	where $\varepsilon_{\mathrm{A}}$, $\varepsilon_{\mathrm{B}}$, $\varepsilon_{\mathrm{m}}$ are the frequency-dependent
	dielectric responses for \textit{A}, \textit{B}, and \textit{m}, respectively.
Accordingly, the experiments demonstrating long-range Casimir
	repulsion were majorly carried out in high-refractive-index fluids,
	i.e., \textit{m} = fluid, in which $\varepsilon_{\mathrm{m}}$ is between $\varepsilon_{\mathrm{A}}$
	and $\varepsilon_{\mathrm{B}}$ over a wide range of frequencies to obey
	inequality \eqref{eq:eps-ineq} \cite{Munday_2009_afm,Milling_1996_afm,Meurk_1997_afm,Lee_2002_afm}.

	Unfortunately, the fluid dielectric response usually drops rapidly beyond the visible frequency region, lowering $\varepsilon_{\mathrm{m}}$ below $\varepsilon_{\mathrm{A}}$ and $\varepsilon_{\mathrm{B}}$ that results in high-frequency attraction\cite{Bostrom_2012_rep}. The long-range repulsive force observed in fluid arises from the retardation effect that diminishes the high-frequency contributions, but when working at small separations, the full-spectrum summation may convert the force from repulsion to attraction \cite{Bostrom_2012_h2_gr}.
	As the London dispersion remains an important part of the interactions, the measured Casimir-Lifshitz repulsion was rather weak (in the order of 0.1 - 10 N/m$^2$)\cite{Munday_2009_afm}.
	
	\section*{Results and disscussions}
	
	\subsection*{Modeling dielectric response of 2D material}
	
\begin{figure}
		\centering
		\includegraphics[scale=1]{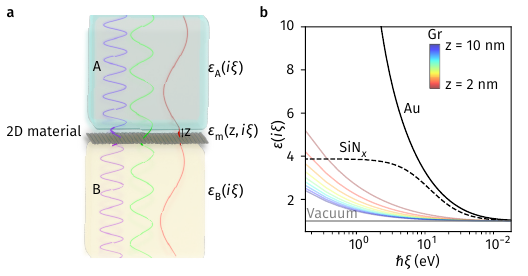}
		\caption{\textbf{Lifshitz-vdW repulsion at surfaces of suspended 2D materials.} 
			\textbf{(a)} The interaction potential between materials \textit{A} and \textit{B} across a birefringent medium gap \textit{m} containing a sheet of monolayer 2D material becomes repulsive when $\left[\varepsilon_{\mathrm{A}}(i\xi)-\hat{\varepsilon}(i\xi)\right]\left[\varepsilon_{\mathrm{B}}(i\xi)-\hat{\varepsilon}(i\xi)\right]<0$. 
			\textbf{(b)} Dielectric responses for Au, SiN$_{\mathrm{x}}$, Vacuum and Gr as a function of electromagnetic energy, $\hbar\xi$, for different separations. Accordingly, we predict that Lifshitz-vdW repulsion may be observed for A/m/B = Vac/Gr/Au at any separation for all frequencies.}
		\label{fig:1}
	\end{figure}

	Inspired by recent findings of the wetting transparency \cite{Rafiee_2012_trans,Shih_2012_prl,Li_2018_vdw,Liu_2018_screening_ml,Ambrosetti_2018_carbon} and the remote epitaxy on graphene-coated substrates \cite{Kim_2017_remote,Kong_2018_pol} and suspended graphene \cite{Frances_2022},
	we conceived a material system generating Lifshitz repulsion, in which metal (\textit{B}) interacts with vacuum (\textit{A}) across a layer of suspended 2D material (\textit{m})  (Fig. \ref{fig:1}a).
	When the separation $z$ is larger than the monolayer thickness, we treat the separation gap containing vacuum stacked on the monolayer as an effective birefringent medium with in-plane (IP) and out-of-plane (OP) dielectric responses $\varepsilon^{\parallel}_{\mathrm{m}}$ and $\varepsilon^{\perp}_{\mathrm{m}}$, which are functions of $z$ and imaginary frequency $i\xi$, resulting from distinct IP and OP electronic properties of the monolayer.
	Indeed, from a dielectric screening point of view, recent findings according to the density functional theory (DFT) calculations have suggested that the dielectric response for a sheet of suspended graphene is highly influenced by the size of surrounding vacuum \cite{Tian_2019_nanolett}. In order to properly model the birefringence for the medium taking into account the surrounding vacuum, based on recent Lifshitz formalism for the calculation of vdW interactions of layered material \cite{Tian_2019_nanolett, Klimchitskaya_2022,Zhou_2018_lifshitz2}, 
	it follows $\varepsilon^{\parallel}_{\mathrm{m}}$ and $\varepsilon^{\perp}_{\mathrm{m}}$ are given by $\varepsilon_{\mathrm{m}}^{\parallel}(z) = 1 + \dfrac{\alpha_{\mathrm{2D}}^{\parallel}}{\varepsilon_{0} z}$ and $\varepsilon_{\mathrm{m}}^{\perp}(z) = \left(1 - \dfrac{\alpha_{\mathrm{2D}}^{\perp}}{\varepsilon_{0} z} \right)^{-1}$, respectively, where $\alpha_{\mathrm{2D}}^\parallel$ and $\alpha_{\mathrm{2D}}^\perp$ are the $z$-independent IP and OP polarizabilities for the 2D material extracted from first-principles calculation (see Materials and Methods in Supplementary Material).
	
	Figure \ref{fig:1}b compares the dielectric responses for gold (Au), silicon nitride (SiN$_{\mathrm{x}}$), vacuum (Vac), and graphene (Gr) for different separations $z$. For graphene’s dielectric responses, the geometrically-averaged dielectric functions $\hat{\varepsilon}_{\mathrm{m}} = \sqrt{\varepsilon_{\mathrm{m}}^\parallel \varepsilon_{\mathrm{m}}^\perp }$ are used here.
	Our calculations reveal that the vdW repulsive forces may be generated in two sets of material systems: (i) A/m/B = Vac/Gr/Au and (ii) A/m/B = Vac/Gr/SiNx. In particular, the former obeys the dielectric mismatch condition in inequality \eqref{eq:eps-ineq} in all separations and frequencies, yielding full-spectrum repulsion.

	We note that the earliest demonstration of Lifshitz-vdW repulsion shared similar scenario, in which the repulsion generated between container wall (\textit{B}) and vacuum (\textit{A}) though a superfluid helium film (\textit{m}) resulted in fluid climbing\cite{Sabisky_1973_liq_He}. However, in our case, the replacement of fluid films with 2D monolayers not only enables significantly wider spectral coverage, but also permits sub-10 nanometer separations, which enhance the interaction, as the energy defined by the Lifshitz formulism scales with the inverse-square law within the vdW regime\cite{Parsegian_2010_vdW,Buhmann_2012,Buhmann_II_2012}.  
	
	\subsection*{Direct measurement of vdW repulsion}
	
\begin{figure}
		\centering
		\includegraphics[scale=1]{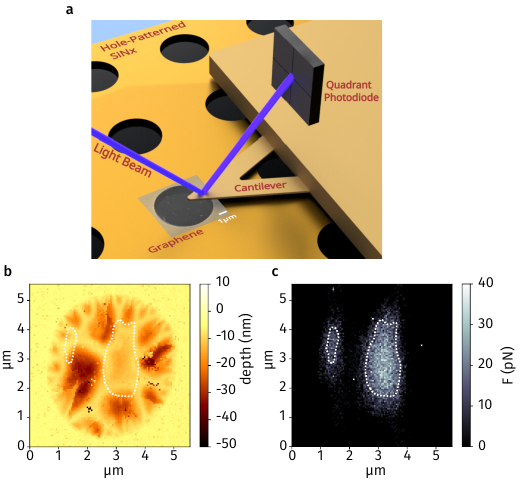}
		\caption{
\textbf{Direct measurement of the Lifshitz-vdW repulsion on suspended graphene.} 
				\textbf{(a)} Schematic diagram for the AFM measurement of the interaction forces experienced by a gold-coated AFM tip approaching suspended and SiN$_{\mathrm{x}}$-supported graphene.
				\textbf{(b),(c)} Representative topographical \textbf{(b)} and corresponding surface force \textbf{(c)} maps for a 33 nm diameter gold-coated AFM tip interacting with a sheet of micromechanically exfoliated graphene transferred on to a SiN$_{\mathrm{x}}$ holey substrate, suspending on a 5 $\mathrm{\mu}$m-diameter hole. The surface force map presents the vertical force values experienced by the AFM tip at a displacement $d$ of approximately 10 nm before establishing the contact. The area enclosed by white dots corresponds to continuous regions experiencing repulsive forces of $\geq$ 10 pN.
}
		\label{fig:2}
	\end{figure}

	Motivated by the analysis presented priorly, we firstly validate our hypothesis by direct measurement of surface forces on a sheet of suspended graphene using atomic force microscopy (AFM). A gold-coated silicon nitride (SiN$_{\mathrm{x}}$) tip with a measured radius of 33 nm was chosen for the force-displacement measurements (see Section S1 in Supplementary Material). 
	The suspended graphene was fabricated by transferring a piece of mechanically-exfoliated graphene to a holey SiN$_{\mathrm{x}}$ membrane\cite{doi:10.1021/nl1008037}, with a hole diameter of approximately 5 $\mathrm{\mu}$m, followed by annealing it in Ar/H$_2$ to remove contaminants \cite{Li_2013_contam,Russo_2014_h2}. A schematic diagram of the measurement system is shown in Figure \ref{fig:2}a. Upon AFM tip displacement, the experienced attractive or repulsive forces were recorded until establishing contact, which we define as the reference point corresponding to tip displacement, $d=0$ nm, at minimum force in each measurement.
	
	One noteworthy observation on the first measurements was that when establishing the contact, the force response is quadratic for freestanding graphene, in contrast to the linear response on supported region (see Supplementary Figures S4). This is well-known considering the mechanical flexibility of freestanding graphene membrane, which yields elastic response of higher order \cite{doi:10.1126/science.1157996}. Indeed, during the retraction process from a freestanding graphene surface, the tip remains to adhere to graphene at a large tip displacement, revealing that both graphene and AFM cantilever were bent before breaking the physical contact. We also observed that the required force to break the physical contact is approximately equivalent for both suspended and supported graphene. We therefore infer that the dominating component of the contact mechanism is caused by capillary or meniscus forces\cite{Meyer_1992,Cappella_1999}, due to the condensation of water within the small gap between the tip and graphene surfaces. With the nonideality in mind, hereafter, we focus on the approach responses before physical contact with the sample surface.

	\begin{figure}
		\centering
		\includegraphics[scale=1]{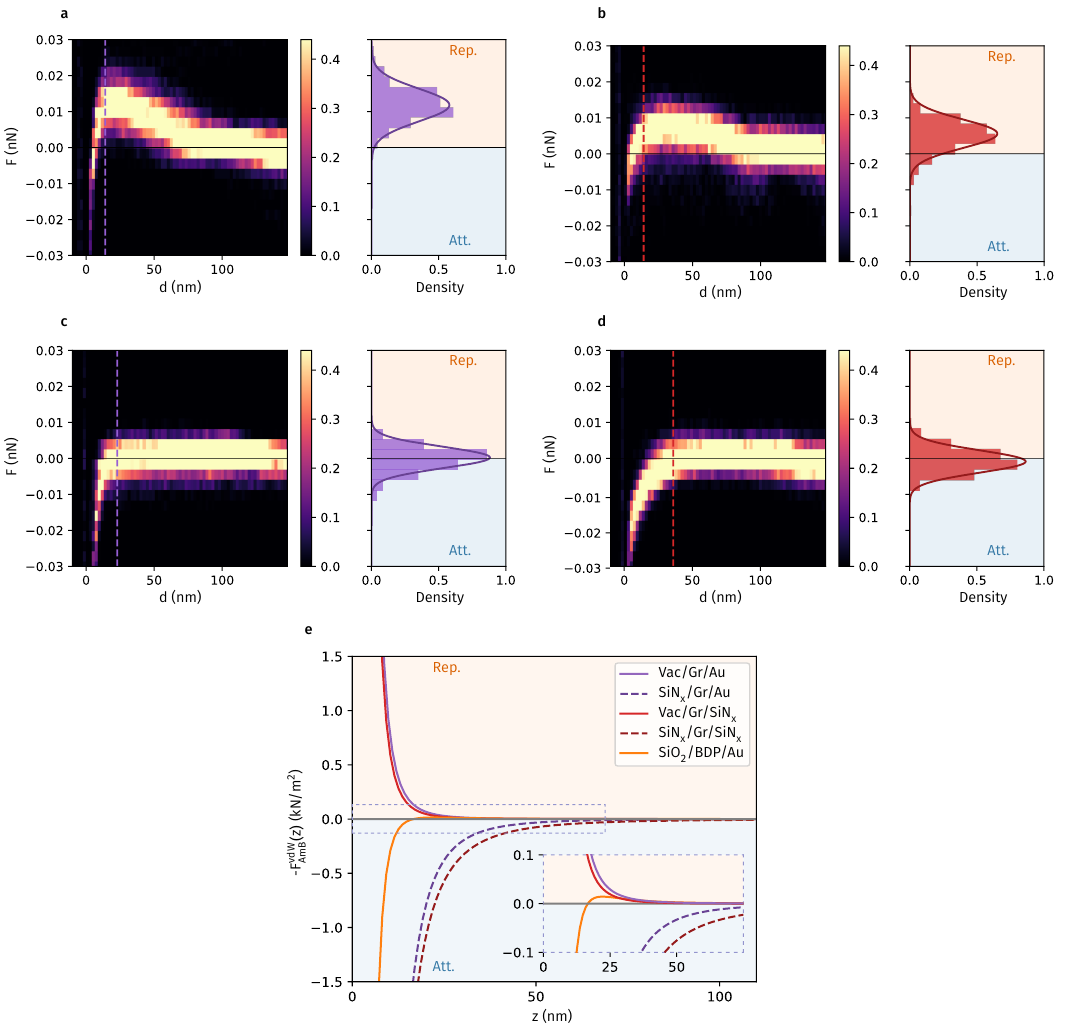}
		\caption{\textbf{Comparison of Lifshitz-vdW force-displacement responses in different systems.} 2D histograms of force-displacement responses for a gold-coated and an uncoated SiN$_{\mathrm{x}}$ AFM approaching \textbf{(a)},\textbf{(b)} flat suspended graphene and \textbf{(c)},\textbf{(d)} SiN$_{\mathrm{x}}$-supported graphene, respectively. \textbf{(e)} Comparison of Lifshitz theory-calculated vdW forces per unit area as a function of separation $z$ for A/m/B = Vac/Gr/Au, Vac/Gr/SiN$_{\mathrm{x}}$, SiN$_{\mathrm{x}}$/Gr/Au, SiN$_{\mathrm{x}}$/Gr/SiN$_{\mathrm{x}}$ and SiO$_2$/BB/Au.
		} 
		\label{fig:3}
	\end{figure}
	
	Based on 12,321 independent force-displacement measurements scanning over an $5.5 \times 5.5$ $\mathrm{\mu}$m$^2$ area of monolayer graphene transferred on a SiN$_{\mathrm{x}}$ hole (scanning electron microscope (SEM) image see Fig. \ref{fig:2}a), Figs. \ref{fig:2}b and \ref{fig:2}c present the topographical and force maps, respectively, showing the surface force experienced by the AFM tip at $d=$ 10 nm. We first noticed that the area of suspended graphene is not perfectly flat, with a degree of surface corrugation created near the SiN$_{\mathrm{x}}$ hole edge. Indeed, as the transfer process involves liquid drying for robust adhesion, the dissipation of liquid surface tension would unavoidably lead to some strain upon suspension. Nevertheless, there remains an atomically-flat and contamination-free area of greater than  5 $\mathrm{\mu}$m$^2$ near the center of suspended graphene. This region of suspended graphene is mechanically flat and held by SiN$_{\mathrm{x}}$ membrane.

	We found the surface forces detected on the corrugated graphene for $d \geq$ 10 nm are nearly negligible (see Section S2 in Supplementary Material). This is unsurprising because a large degree of interaction is internally absorbed by the soft nature of corrugated graphene. On the other hand, remarkably, the AFM tip consistently experienced repulsive forces on flat (or stretched) suspended graphene. The repulsion observed here cannot result from charge interactions because both graphene and gold are conductive; any charge trapped on graphene surface will induce an image charge of opposite sign in gold that only leads to an attractive interaction. In addition, the repulsive area is electrically connected to corrugated and SiN$_{\mathrm{x}}$-supported graphene, such that any residual charge will be neutralized over the whole area rather than localization.

	Figures \ref{fig:3}a-\ref{fig:3}d compare 2D histograms for 144 force-displacement responses extracted from independent measurements of scanning force microscopy over an $600 \times 600$ nm$^2$ area of a sheet of suspended and SiN$_{\mathrm{x}}$-supported graphene, using a gold-coated (Figs. \ref{fig:3}a and \ref{fig:3}c) and uncoated (Figs. \ref{fig:3}b and \ref{fig:3}d) SiN$_{\mathrm{x}}$ tip of radius of 20 nm. 
	The right panels present the force distributions at given $d$-cuts associated with the dashed lines in the left panels.
	For the measurements on suspended graphene, both gold-coated and uncoated AFM tip started to experience repulsion from $d <$ 75 nm, followed by a gradual increase with decreasing displacement. 
The last notable measured repulsive force were detected at an average displacement  8.8 nm and 6.6 nm for gold and SiN$_{\mathrm{x}}$ AFM tips, respectively. The corresponding Gaussian fits reveal a mean repulsive force of 11.8 $\pm$ 4.6 and 5.7 $\pm$ 4.0 pN for gold and SiN$_{\mathrm{x}}$ tips experienced on suspended graphene, respectively. When the AFM tip further approached the suspended graphene, we observed a sudden emergence of attractive force, hypothetically resulting from the earlier mentioned capillary or meniscus forces \cite{Meyer_1992,Cappella_1999}.
	 	 
	On the other hand, only attractive responses were recorded on SiN$_{\mathrm{x}}$-supported graphene (Fig. \ref{fig:3}c and \ref{fig:3}d) for both tips.
The interaction appears to be relatively short-range, nearly negligible for large separations above 40 nm. The gold-coated and uncoated SiNx tips experienced weak attractive forces of -0.6 pN $\pm$ 3.4 pN and -2.0 $\pm$ 2.9  pN at $d=$ 18.5 nm and $d=$ 36.0 nm, respectively (Figs. \ref{fig:3}c and \ref{fig:3}d right), corresponding to the onset of attraction.
At smaller displacements, we observed monotonically attractive forces following typical force-displacement dependency for an AFM tip approaching an electroneutral solid surface \cite{Cappella_1999,Voigtlnder_2019}. 
	
	In order to examine the measured force-displacement responses on graphene surfaces, we have modeled the vdW interactions using the Lifshitz theory. Specifically, the vdW interaction between two bulk materials \textit{A} and \textit{B} per unit area, across a birefringent medium \textit{m}, $\Phi_{\mathrm{AmB}}^{\mathrm{vdW}}$, as a function of separation $z$, is given by \cite{Parsegian_2010_vdW}:
	
	\begin{equation}
		\label{eq:Phi-aniso}
		\Phi_{\mathrm{AmB}}^{\mathrm{vdW}} (z)
= \sum_{n = -\infty}^{\infty}
		\frac{k_{\mathrm{B}} T g_{\mathrm{m}}(i \xi_{n})}{16 \pi z^{2}} 
		\left\{
		\int_{\mathcal{r}_{n}}^{\infty} q \ln
		\left[
		1 - \Delta_{\mathrm{Am}}(i \xi_{n}) \Delta_{\mathrm{Bm}}(i \xi_{n}) e^{-q}
		\right] \mathrm{d} \mathcal{q}
		\right\}
	\end{equation}
where $k_{\mathrm{B}}$ is the Boltzmann constant, $T$ is the absolute
	temperature, $\xi_{n}=2\pi n k_{\mathrm{B}} T/ \hbar$ is the n-th
	Matsubara frequency, $\hbar$ is the reduced Planck constant,
	$\mathcal{r}_{n}=\frac{2d \xi_{n}}{c} \sqrt{\hat{\varepsilon}_{\mathrm{m}}}$ is
	the retardation factor\cite{Parsegian_2010_vdW}, $c$ is the speed
	of light in vacuum and $q$ is a dimensionless auxiliary variable.
	$g_{\mathrm{m}} = \varepsilon_{\mathrm{m}}^\perp / \varepsilon_{\mathrm{m}}^\parallel$ is the dielectric
	anisotropy\cite{Tian_2019_nanolett} of m.
Note that this approach was also used to calculate the vdW interactions of layered materials \cite{Zhou_2017_lifshitz}, using the effective medium approach to compute the dielectric responses of monolayers in vacuum, as we illustrated
	earlier.
$\Delta_{\mathrm{Am}}$ and $\Delta_{\mathrm{Bm}}$ correspond to the dielectric
	mismatches following
	$ \Delta_{\mathrm{jm}} = \dfrac{ \hat{\varepsilon}_{\mathrm{j}} -
		\hat{\varepsilon}_{\mathrm{m}} }{ \hat{\varepsilon}_{\mathrm{j}} + \hat{\varepsilon}_{\mathrm{m}}}$,
	for j = \textit{A}, \textit{B}. Analogous to inequality \ref{eq:eps-ineq}, the
	vdW potential for a given EM mode $\xi_{n}$ becomes
	positive when $\Delta_{\mathrm{Am}} \Delta_{\mathrm{Bm}} < 0$, contributing to
	vdW repulsion. 
	Accordingly, the vdW force per unit area generated between \textit{A} and \textit{B}, $F^{\mathrm{vdW}}_{\mathrm{AmB}}$, is given by\cite{Dzyaloshinskii_1961_lifshitz,Buhmann_2012,Buhmann_II_2012}: 
	\begin{equation}
		\label{eq:F-aniso}
		F_{\mathrm{AmB}}^{\mathrm{vdW}} (z)
=
		\sum_{n = -\infty}^{\infty}
		\frac{k_{\mathrm{B}} T  g_{\mathrm{m}}(i \xi_{n})}{16 \pi z^{3}} 
		\left\{
		\int_{r_{n}}^{\infty} q^2 \frac{\Delta_{\mathrm{Am}}(i \xi_{n}) \Delta_{\mathrm{Bm}}(i \xi_{n}) e^{-q}}{
			1 - \Delta_{\mathrm{Am}}(i \xi_{n}) \Delta_{\mathrm{Bm}}(i \xi_{n}) e^{-q}}
		\mathrm{d} q
		\right\}
	\end{equation}
	Although the dielectric response for 2D material is $z$-dependent, namely $	\hat{\varepsilon}_{\mathrm{m}}(z,\xi_{n})$, its partial derivative with respect to $z$ only contributes to correction of high orders. Eq. \eqref{eq:F-aniso} is sufficiently accurate to approximate the exact solution. 
	
	Figure \ref{fig:3}e presents the calculated $-F^{\mathrm{vdW}}_{\mathrm{AmB}}$, as a function of separation $z$ for A/m/B = Vac/Gr/Au, Vac/Gr/SiN$_{\mathrm{x}}$, SiN$_{\mathrm{x}}$/Gr/Au and SiN$_{\mathrm{x}}$/Gr/SiN$_{\mathrm{x}}$. We further compare the calculated response for A/m/B = SiO$_2$/Bromobenzene(BB)/Au, benchmarking the fluid immersion system that quantitatively demonstrates Lifshitz-Casimir repulsion at large separations \cite{Munday_2009_afm}.
The three systems showing repulsive responses, A/m/B = Vac/Gr/Au, Vac/Gr/SiN$_{\mathrm{x}}$ and SiO$_2$/BB/Au, have similar strength for $z>$ 30 nm.
	The liquid immersion system is even slightly larger for $z >$ 50 nm (See Fig.S2 in Supplementary Material), as graphene’s dielectric response drops rapidly with increasing separation. Nevertheless, as pointed out by Boström et al \cite{Bostrom_2012_rep}, in SiO$_2$/BB/Au system, the receding retardation effect turns the Casimir repulsion to vdW attraction for $z <$ 20 nm, exhibiting a maximum repulsive force of $\sim$0.015 kN/m$^2$ at $z \approx 25$ nm.
	
	Remarkably, both Vac/Gr/Au and Vac/Gr/SiN$_{\mathrm{x}}$ systems remain repulsive for small separations. As the repulsion scales approximately with $z^{-3}$, the Lifshitz theory predicts strong repulsive forces  at separations of 9 nm, respectively of $\sim$1.3 kN/m$^2$ for gold and $\sim$0.9 kN/m$^2$ for SiN$_{\mathrm{x}}$, which are orders-of-magnitude higher than the fluid immersion system. In both systems, we notice that the repulsive forces are approaching atmospheric pressure at even smaller separations.
	
	For the interactions with SiN$_{\mathrm{x}}$-supported graphene, the calculated $-F^{\mathrm{vdW}}_{\mathrm{AmB}}(z)$ for gold and SiN$_{\mathrm{x}}$ (Fig. \ref{fig:3}c and \ref{fig:3}d) nicely capture the features of the measured $F-d$ responses, in which the attractive force starts to emerge at $d$ (or separation $z$) of $\approx$ 30 nm (Fig. \ref{fig:3}e). In addition, the experimentally observed attractive interaction with SiN$_{\mathrm{x}}$ is more long-ranged compared to gold, in coherence with the calculations shown in Fig. \ref{fig:3}e. 
	
	However, we realized that it is not proper to directly evaluate the AFM-measured $F-d$ responses on suspended graphene using the calculated $-F^{\mathrm{vdW}}_{\mathrm{AmB}}(z)$ profile. A major concern is that upon AFM indentation, the displacement of AFM tip, $d$, is not equal to the separation between Au and Vac, $z$, owing to the mechanical deformation of suspended graphene. More specifically, when AFM tip is approaching, the generated repulsion rather presses and deforms the suspended graphene, so that the tip displacement does not effectively reduce the separation between tip and graphene. The scenario explains why the repulsion was detected already at large displacements ($d \approx$ 75 nm; see Fig. \ref{fig:3}a and Fig. \ref{fig:3}b), as compared to the theoretically predicted onset separation ($z \approx$ 20 nm; see Fig. \ref{fig:3}e). 
	
\begin{figure}
		\centering
		\includegraphics[scale=1]{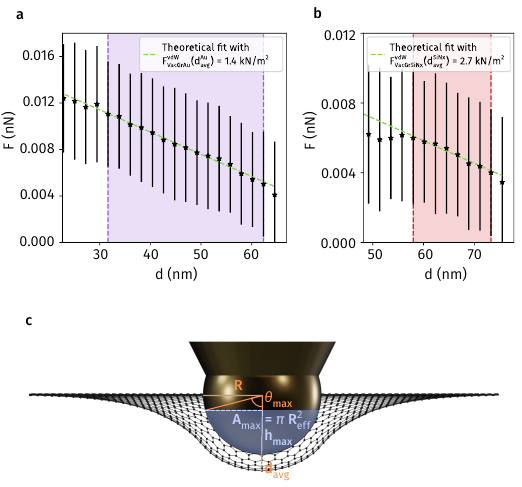}
		\caption{\textbf{Comparison of measured repulsive forces with Lifshitz theory taking into account of deformation of suspended graphene.} 
		AFM-measured repulsive forces with standard deviations as a function of $d$ for gold \textbf{(a)} and SiN$_{\mathrm{x}}$ \textbf{(b)} AFM tips. There exists a regime where repulsive force linearly increases with $d$, where the tip displacement only deforms graphene without changing the average separation between tip and graphene, $d_{\mathrm{avg}}$. 
		\textbf{(c)} Schematic diagram showing the deformation of suspended graphene and the extracted geometric parameters at $d = d_{\mathrm{avg}}$, yielding the maximum repulsive force detected between tip and graphene.
		} 
		\label{fig:4}
	\end{figure}

	We have developed a model taking into account the effect of deformation of suspended graphene. 
Fig. \ref{fig:4}a and Fig. \ref{fig:4}b present magnified $F-d$ responses together with standard deviations as a function of $d$ extracted from Fig. \ref{fig:3}a and \ref{fig:3}b. We found that there exists linear regimes, $25 < d < 62$ nm for gold (Fig. \ref{fig:4}a) and $58 < d < 73$ nm for SiN$_{\mathrm{x}}$ (Fig. \ref{fig:4}b) tips, where the detected repulsive forces increase linearly by reducing the displacement. One could notice that the width of these regimes are approximately equal to the corresponding AFM tip radii, 33 nm and 20 nm for gold and SiN$_{\mathrm{x}}$, respectively. We therefore infer that the linear regime essentially informs the repulsive force experienced by the AFM tip during an indentation process that deforms graphene from a plat plain to a hemispherical surface, as revealed in Fig. \ref{fig:4}c, without reducing the average separation between the tip and graphene, $d_{\mathrm{avg}}$.
	
	More specifically, within the linear regime, when the AFM tip of a hemispherical tip with radius of $R$ approaches suspended graphene, the repulsive force is sufficiently strong, such that the increase of measured force with displacement primarily results from the increase of interacting area, $A$. Indeed, considering the facts of (i) relatively small dielectric response $\varepsilon$ of SiN$_{\mathrm{x}}$ compared to Au (Fig.\ref{fig:1}b) and (ii) similar cantilever stiffness, $k_{\mathrm{Au}}$ = 0.16136 N/m  and  $k_{\mathrm{SiN}_{\mathrm{x}}}$ = 0.11670 N/m, it is reasonable to infer that the average separation required to bend the AFM cantilever for the SiN$_{\mathrm{x}}$ tip is smaller than the gold counterpart, $d^{\text{SiN}_{\mathrm{x}}}_{\mathrm{avg}} < d^{\text{Au}}_{\mathrm{avg}}$. With this in mind, given the significant deflection of repulsive force at an average displacement $d =$ 8.8 nm for gold and $d =$ 6.6 nm for SiN$_{\mathrm{x}}$ before bouncing into contact, we let $d^{\text{Au}}_{\mathrm{avg}}=$ 8.8 nm and $d^{\text{SiN}_{\mathrm{x}}}_{\mathrm{avg}}=$ 6.6 nm for proper comparison with our calculations. Following the scenario, beyond the linear regime, the repulsive force exhibited a plateau for 8.8 $< d <$ 25 nm and 6.6 $< d <$ 58 nm, for gold and SiN$_{\mathrm{x}}$ tips, respectively, where the interacting area between AFM tip and deformed graphene remained nearly unchanged upon indentation, reaching its maximum, $A_{\mathrm{max}}$ (see Fig. \ref{fig:4}c).  
	
	According to the physical picture presented above, considering the spherical geometry of the tip, we model the slope of the $F-d$ response in the linear regime, $m_{\mathrm{fit}}$, following:
	
	\begin{equation}
		\label{eq:Reff}
		m_{\mathrm{fit}} = - F_{\mathrm{AmB}}^{\mathrm{vdW}} (z = d_{\mathrm{avg}}) \pi R_{\text{eff}}
	\end{equation}

	where $R_{\mathrm{eff}}$ is the effective radius of the projected circular area interacting with graphene (see Fig. \ref{fig:4}c). Using the experimentally extracted value $m^{\text{Au}}_{\mathrm{fit}} =$ -0.20 mN/m and $m^{\text{SiN}_{\mathrm{x}}}_{\mathrm{fit}} =$ -0.13 mN/m and the Lifshitz-theory calculated force density  $F_{\mathrm{VacGrAu}}^{\mathrm{vdW}} (z = d^{\text{Au}}_{\mathrm{avg}})$ = 1.4 kN/m$^2$ and $F_{\mathrm{VacGrSiN}_{\mathrm{x}}}^{\mathrm{vdW}} (z = d^{\text{SiN}_{\mathrm{x}}}_{\mathrm{avg}})$ = 2.7 kN/m$^2$, the geometrical parameters $R_{\mathrm{eff}}$ and $A_{\mathrm{max}}$ at the deflection points $d^{\text{Au}}_{\mathrm{avg}}$ = 8.8 nm and $d^{\text{SiN}_{\mathrm{x}}}_{\mathrm{avg}}$ = 6.6 nm  (see Fig. \ref{fig:4}c) are determined to be 43 nm and 5949 nm$^2$ for gold and 15 nm and 720 nm$^2$ for SiN$_{\mathrm{x}}$, respectively. Remarkably, all extracted parameters are self-consistent with the physical picture proposed here and match the geometrical constraints of the measurement procedure. 
	
	\section*{Nonwettability of suspended graphene}
	
	Our findings imply that suspended 2D monolayers are intrinsically repulsive surfaces hindering molecular adsorption and deposition, with substantially lowered surface wettability. Indeed, the polarizability theory of 2D materials \cite{Tian_2019_nanolett} suggest that the effective dielectric constant of a sheet of suspended 2D monolayer is only slightly higher than vacuum, particularly at high frequencies and large separations. As a result, the Lifshitz theory suggests that any electroneutral object would experience repulsive forces, smaller or larger, when brought in close proximity on suspended 2D material surfaces.
	
	\begin{figure}[!htbp]
		\centering
		\includegraphics[scale=1]{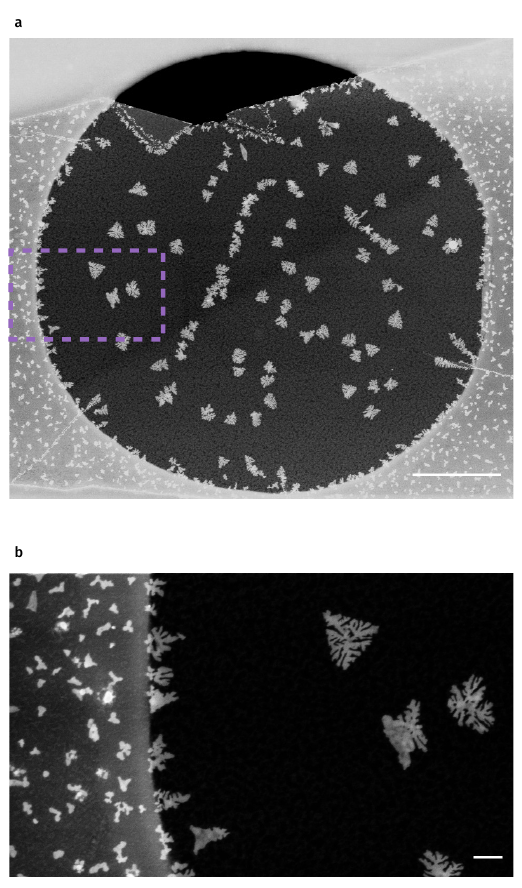}
		\caption{\textbf{Nonwettability of suspended graphene to the nucleation of gold vapor particles.}\textbf{(a)} On SiN$_{\mathrm{x}}$-supported graphene, fast condensation at room temperature yields small nanoclusters with a high nucleation density, whereas on suspended graphene, the nucleation density is extremely low down to $\approx$ 4.6 $\mathrm{\mu}$m$^{-2}$. SEM image scale bar: 1 $\mathrm{\mu}$m. (b) An magnified image for the interface between the SiN$_{\mathrm{x}}$-supported and suspended graphene regions. Individual nuclei within the suspended region grew to form large and dendritic structures. SEM image scale bar: 100 nm.}
		\label{fig:5}
	\end{figure}
	
We examined the postulate by depositing gold particles on suspended graphene through an electron-beam evaporation process in high vacuum. The evaporation source generated high-energy gold vapor particles, which then condensed on a sheet of exfoliated graphene transferred onto a SiN$_{\mathrm{x}}$ window. We deposited a small amount of gold ($\approx$ 1.93 ng/mm$^{2}$) at a rate of 0.01 nm/s on the sample surface at room temperature. As revealed in the SEM images of Fig. \ref{fig:5}, one can clearly identify two regions, namely the SiN$_{\mathrm{x}}$-supported and suspended graphene, which exhibit distinct morphology and density for the deposited gold clusters. On SiN$_{\mathrm{x}}$-supported graphene, due to very high surface energy of gold, fast condensation at room temperature yields small nanoclusters with a high nucleation density. However, on suspended graphene, despite a high degree of supercooling, the nucleation density is extremely low down to $\approx$ 4.6 $\mathrm{\mu}$m$^{-2}$ , with a large non-wettable area of several $\mathrm{\mu}$m$^2$. Individual nuclei within the suspended window grew to form large and dendritic structures (Fig. \ref{fig:5}b).
	
	The extremely low wettability observed here can neither result from the atomic smoothness of graphene nor from any ultra high vacuum treatment, because the nucleation density of gold on single crystalline graphite cleaved in air remains high \cite{Wayman_1975_depo_au_gr1}. Indeed, the classical nucleation theory suggests that the nucleation density is proportional to $D^{-1/3}$, where $D$ is the surface diffusivity \cite{Mo_1991_diffuse}. The extremely low nucleation density essentially suggest that the in-plane movement is nearly frictionless; nearly all gold particles moved to the SiNx-supported region. In fact, we notice that a recent literature by Frances et al \cite{Frances_2022} pointed out the nucleation of gold vapor on suspended graphene takes place exclusively at defects and contaminants. Accordingly, we could conclude that suspended graphene is unwettable to gold particles. The physical picture of strong many-body vdW repulsion formed in Vac/Gr/Au system explains the observations.
	
	\section*{Conclusion}
	We have theoretically and experimentally demonstrated that a strong repulsive Lifshitz vdW forces can be generated on suspended graphene due to sub-10 nanometer separations and full-spectrum repulsion enabled by suspended 2D monolayer, thereby substantially reducing the surface wettability during an epitaxial evaporation process. Future experimental and theoretical investigations could further strengthen our findings and validate implications suggested here. The generation of strong Lifshitz-vdW repulsion can be used to realize quantum levitation \cite{Munday_2009_afm,Zhao_2019_casimir_trap,Munday_2010_review} without fluid immersion, which give rise to new nanoelectromechanical systems. In general, we believe that the manipulation of surface forces and processing of suspended 2D materials will be greatly facilitated by the fundamental insights presented here.
	
	\pagebreak
	\section*{Author Contributions}
	\label{sec:author-contributions}
	G.V., T.T.  and C.J.S. conceived the idea and designed the
	experiments. G.V., T.T. and F.N. developed the theoretical
	framework. T.T. performed first-principle calculations under guidance
	of E.J.G.S.. 
	G.V. of carried out AFM force measurement, analyzed the data, and modeled the force responses with the help of H.J.. 
	G.V., T.T. and K.C. fabricated the freestanding graphene
	samples. G.V., T.T. characterized the freestanding graphene samples.
	G.V., T.T. and C.J.S. co-wrote the paper. All authors contributed to this work, read
	the manuscript, discussed the results, and agreed to the contents of
	the manuscript and supplementary materials.
	
	\section*{Acknowledgements}
	\label{sec:acknowledgments}
	C.J.S. is grateful for financial support
	from ETH startup funding and the European Research Council Starting
	Grant (N849229 CQWLED).
E.J.G.S. acknowledges computational resources through CIRRUS Tier-2 HPC Service (ec131 Cirrus Project) at EPCC (http://www.cirrus.ac.uk) funded by the University of Edinburgh and EPSRC (EP/P020267/1); ARCHER UK National Supercomputing Service (http://www.archer.ac.uk) \textit{via} Project d429. E.J.G.S. acknowledges the EPSRC Open Fellowship (EP/T021578/1), and the Edinburgh-Rice Strategic Collaboration Awards for funding support. 

\pagebreak

	\section*{Supplementary materials}
	\label{sec:suppl}
	\begin{itemize}
		\item Materials and Methods
		\item Supplementary Text
		\item Figs. S1 to S7
\item References (S1-S15)
	\end{itemize}
	
\pagebreak	
	
\printbibliography{}



\section*{Supplementary material}

## Materials and Methods

**Calculation of vdW interaction spectra**   The dielectric response as a function of imaginary frequency $\varepsilon(i\xi)$ is calculated using the Kramers Kronig relationship[S1] given by:

$$\varepsilon(i\xi) = 1 + \frac{2}{\pi} \int_0^\infty \frac{\omega \mathrm{Im}[\varepsilon(\omega)]}{\omega^2 + \xi^2} d\omega \tag{S1}$$

where $\omega$ is the real frequency, and $\mathrm{Im}[\varepsilon(\omega)]$ is the imaginary part of complex dielectric function $\varepsilon(\omega)$. Frequency-dependent dielectric functions of $SiO_2$[S2], $SiN_x$[S2], bromobenzene (BB)[S3] and Au[S2] were extracted from experimental data. Frequency-dependent 2D polarizabilities ($\alpha_{2D}^{\parallel}$, $\alpha_{2D}^{\perp}$) of graphene were calculated using the projector augmented wave method[S4] in the *ab initio* package GPAW[S5]. Dielectric responses were calculated using random phase approximation on top of the Perdew-Burke-Ernzerhof exchange-correlation functional[S6] with plane wave cutoff energy of 500 eV, k-point density of 15 Å$^{-1}$ and truncated Coulomb kernel to avoid spurious interaction from periodic images.

**Exfoliated graphene on silicon nitride chip**   Graphene flakes were exfoliated [S7] from natural graphite and transferred to a silicon nitride ($SiN_x$) chip of 1 cm$^2$ size, patterned with an 12×12 hole-matrix, in which both hole-diameter and separation distance is approximately 5 $\mu$m. The window with the hole-matrix was fabricated using electron beam lithography and etching. A wedging transfer method [S8] was chosen to avoid damaging the hole-patterned structure. During transfer the polymer-graphene structure was aligned with a micro manipulator to the hole matrix.

**Electron microscopy**   SEM characterizations were carried out on Zeiss ULTRA plus with 3 kV beam voltage and 20 $\mu$m aperture.

# S1 Theoretical Simulations

## S1.1 Modified Lifshitz theory for anisotropic media

The derivation of equation (2) is described as follows. The vdW interaction energy $\Phi_{\mathrm{AmB}}^{\mathrm{vdW}}$ corresponding to the total energy summed from all allowed EM modes[S9] is given by:

$$\Phi_{\mathrm{AmB}}^{\mathrm{vdW}} = \frac{k_{\mathrm{B}}T}{2(2\pi)^2} \sum_{n=-\infty}^{\infty} \int_{r_n}^{\infty} \ln \mathcal{D}(i\xi_n, \mathbf{k}) d^2\mathbf{k} \tag{S2}$$

where $\mathbf{k} = (k_x, k_y)$ is the in-plane wavevector, and $\mathcal{D}(i\xi_n, \mathbf{k})$ is the dispersion relation for a given geometry. For generality, the dielectric tensor of material j has diagonal components $\varepsilon_{\mathrm{j}}^{xx}$, $\varepsilon_{\mathrm{j}}^{yy}$ and $\varepsilon_{\mathrm{j}}^{zz}$. By transforming $\mathbf{k} = (\kappa \cos \vartheta, \kappa \sin \vartheta)$, and $g_{\mathrm{j}} = \left[ \frac{\varepsilon_{\mathrm{j}}^{xx}}{\varepsilon_{\mathrm{j}}^{zz}} \cos^2 \vartheta + \frac{\varepsilon_{\mathrm{j}}^{yy}}{\varepsilon_{\mathrm{j}}^{zz}} \sin^2 \vartheta \right]^{-1}$ where $\kappa$, $\vartheta$ are the corresponding polar coordinates of $\mathbf{k}$, the dispersion relation $\mathcal{D}$ of an anisotropic A/m/B layered system follows[S1]:

$$\begin{aligned} \mathcal{D} &= 1 - \underbrace{\left[ \frac{\hat{\varepsilon}_{\mathrm{A}} - \varepsilon_{\mathrm{m}}^{zz} g_{\mathrm{m}}^{1/2}(\vartheta)}{\hat{\varepsilon}_{\mathrm{A}} + \varepsilon_{\mathrm{m}}^{zz} g_{\mathrm{m}}^{1/2}(\vartheta)} \right]}_{\Delta_{\mathrm{Am}}} \\ &\quad \times \underbrace{\left[ \frac{\hat{\varepsilon}_{\mathrm{B}} - \varepsilon_{\mathrm{m}}^{zz} g_{\mathrm{m}}^{1/2}(\vartheta)}{\hat{\varepsilon}_{\mathrm{B}} + \varepsilon_{\mathrm{m}}^{zz} g_{\mathrm{m}}^{1/2}(\vartheta)} \right]}_{\Delta_{\mathrm{Bm}}} e^{-2 g_{\mathrm{m}}^{1/2}(\vartheta) \kappa z} \\ &= 1 - \Delta_{\mathrm{Am}}(\vartheta) \Delta_{\mathrm{Bm}}(\vartheta) e^{-2 g_{\mathrm{m}}^{1/2}(\vartheta) \kappa z} \end{aligned} \tag{S3}$$

By further introducing an auxiliary variable $x = 2 g_{\mathrm{m}}^{1/2}(\theta) \kappa z$, it follows:

$$\Phi_{\mathrm{AmB}}^{\mathrm{vdW}} = \frac{k_{\mathrm{B}}T}{32\pi^2 z^2} \sum_{n=-\infty}^{\infty} \int_0^{2\pi} g_{\mathrm{m}}(i\xi_n, \vartheta) d\vartheta \int_{r_n}^{\infty} x dx \ln[1 - \Delta_{\mathrm{Am}}(i\xi_n, \vartheta) \Delta_{\mathrm{Bm}}(i\xi_n, \vartheta) e^{-x}] \tag{S4}$$

Accordingly, equation (2) is obtained for $\varepsilon_{\mathrm{m}}^{xx} = \varepsilon_{\mathrm{m}}^{yy}$, i.e. $g_{\mathrm{m}}$ is independent of $\vartheta$, which is valid for most 2D materials where the 2D lattice is hexagonal or square. Moreover, in equation (2), $\hat{\varepsilon}_{\mathrm{A}}$ and $\hat{\varepsilon}_{\mathrm{B}}$ reduce to $\varepsilon_{\mathrm{A}}$ and $\varepsilon_{\mathrm{B}}$, respectively, when A and B are isotropic bulk materials. Time-reversal symmetry $\varepsilon(i\xi) = \varepsilon(-i\xi)$ is used when magnetic response of the material is negligible, therefore equation (2) only needs to be evaluated for $\xi_n \geq 0$. Moreover, our numerical analysis suggests the integral of $\Phi_{\mathrm{AmB}}^{\mathrm{vdW}}$ is dominated by $x \leq 5$, or equivalently $\kappa \leq 2.5(g_{\mathrm{m}}d)^{-1}$. When $d$ is in the order of 2 nm, and $g_{\mathrm{m}} = 2.5$, the majority of interaction comes from EM modes with

$\kappa < 0.05$ Å$^{-1}$. In other words, evaluating equation (2) using the material dielectric functions at the optical limit ($\mathbf{k} \to 0$) would preserve the accuracy of $\Phi_{\text{AmB}}$ calculation.

The vdW force per unit area $F_{\text{AmB}}^{\text{vdW}}$ is the derivative of the vdW potenial with respect to $z$ given by:

$$\mathbf{F}_{\text{AmB}}^{\text{vdW}} = -\nabla\Phi_{\text{AmB}} = -\frac{\partial\Phi_{\text{AmB}}^{\text{vdW}}}{\partial z} \cdot \hat{\mathbf{e}}_z \quad (S5)$$
$$= \sum_{n=-\infty}^{\infty} \frac{k_{\text{B}}T g_{\text{m}}(i\xi_n)}{16\pi z^3} \left\{ \int_{r_n}^{\infty} x^2 \frac{\Delta_{\text{Am}}(i\xi_n)\Delta_{\text{Bm}}(i\xi_n)e^{-x}}{1 - \Delta_{\text{Am}}(i\xi_n)\Delta_{\text{Bm}}(i\xi_n)e^{-x}} dx \right\} \cdot \hat{\mathbf{e}}_z$$

which leads us to equation (3) for the vdW force per unit area $F_{\text{AmB}}^{\text{vdW}} = |\mathbf{F}_{\text{AmB}}^{\text{vdW}}|$.

Complementary to Fig. 1b, Fig. S1 compares the dielectric responses of other materials studied here. Notably, at the order of $d = 1$ nm, $\hat{\varepsilon}_{\text{m}}$ of graphene is comparable to that of bromobenzene (BB), a high-refractive-index liquid commonly used in experiments demonstrating Casimir repulsion[S3, S10]. In this respect, 2D material appears to be a promising candidate for realizing repulsive vdW interactions. However, unlike bulk liquid, $\hat{\varepsilon}_{\text{m}}$ of a 2D material strongly depends on the separation $d$, making the repulsion more pronounced at short distances (the vdW regime).

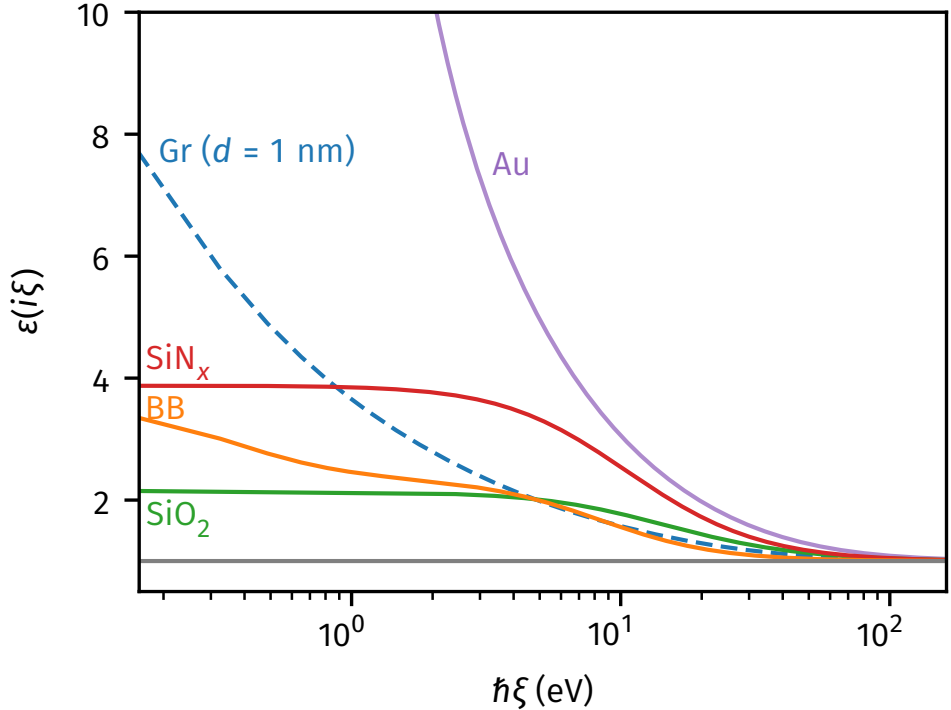

Figure S1. **Comparison of $\hat{\varepsilon}(i\xi)$ responses for materials considered in this study**. The effective dielectric function of graphene at $d = 1$ nm is higher than that of the widely-used high-refractive-index liquid bromobenzene (BB).

We further compared the force response $F_{\text{AmB}}^{\text{vdW}}$ for A/m/B = Vac/Gr/Au, Vac/Gr/SiN$_x$ and SiO$_2$/Bromobenzene(BB)/Au, in which the latter quantitatively demonstrates Lifshitz-Casimir repulsion at large separations [S3]. We can cleary see that the three systems are very representative, generating repulsion in the vdW and Casimir interaction regimes, respectively. Figure S2 shows that for 30 nm, all three interactions are repulsive with similar strength, with the liquid immersion system being slightly larger for $z$ larger than 50 nm, resulting from a reduction of graphene's dielectric response. Nevertheless, as pointed out by Boström et al [S11], in SiO$_2$/BB/Au system, the receding retardation effect turns the Casimir repulsion to vdW attraction for $z < 20$nm, exhibiting a maximum repulsive force of 15 $\frac{\text{N}}{\text{m}^2}$ at approximately 25 nm.

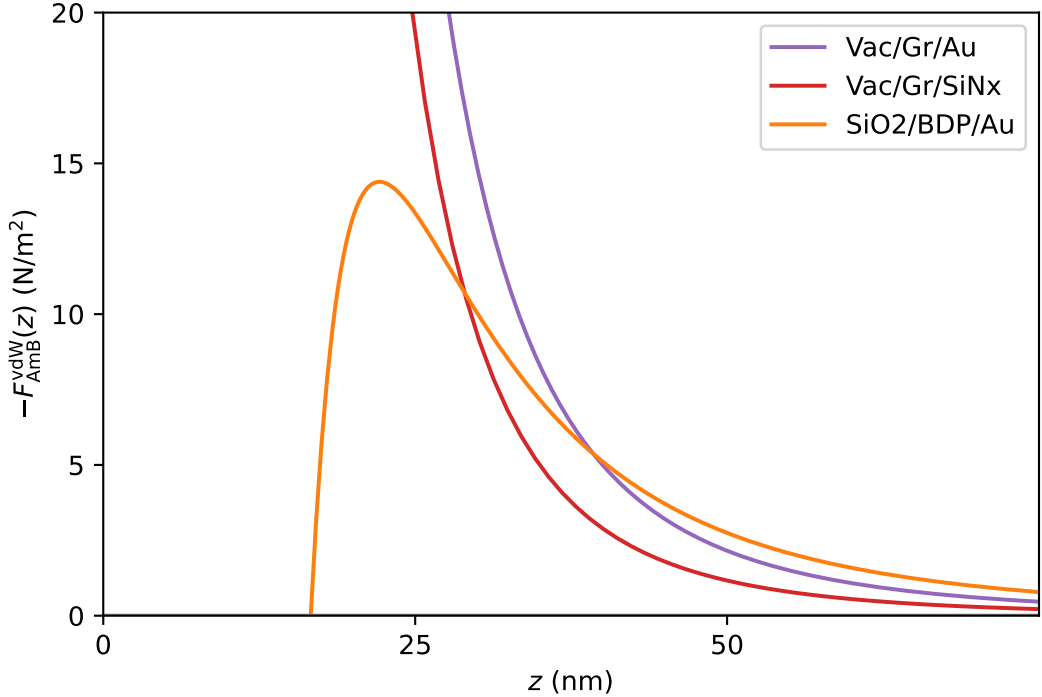

Figure S2. **Comparison of $F_{\text{AmB}}^{\text{vdW}}$ for A/m/B = Vac/Gr/Au, Vac/Gr/SiN$_x$ and SiO$_2$/Bromobenzene(BB)/Au**. For $z > 30$ nm, all three interactions are repulsive with similar strength, with the liquid immersion system being slightly larger for $z > 50$ nm. As pointed out by Boström et al [S11], in SiO$_2$/BB/Au system, the receding retardation effect turns the Casimir repulsion to vdW attraction for small separations $z$.

## S2 Force-distance AFM measurements

### S2.1 AFM setup and calibration

The force-distance measurements were carried out with a Bruker BioScope Resolve AFM using the B side of a gold-coated NPG-10 and an uncoated DNP-10 Bruker (SiN$_x$) AFM tip, with estimated radii $R$ of approximately 33 nm and 20 nm, as highlighted with a red dashed circles in Fig. S3a and S3b, respectively. In order to obtain the interaction force (N) from the original measured values of the piezo element displacement in (mV), we carried out tip calibration steps to extract the deflection sensitivity $\left(\frac{m}{V}\right)$ and spring constant $\left(\frac{N}{m}\right)$, by using the PeakForce™QNM™suite within the NanoScope®software environment. We determined a deflection sensitivity of 69.679 $\frac{nm}{V}$ and a spring constant of 0.16136 $\frac{N}{m}$ for the Au-coated tip and a deflection sensitivity of 75.567 $\frac{nm}{V}$ and a spring constant of 0.11670 $\frac{N}{m}$ for the uncoated tip. The calibrated spring constants are comparable to the nominal value of $0.12\frac{N}{m}$ for both provided by the tip vendor.

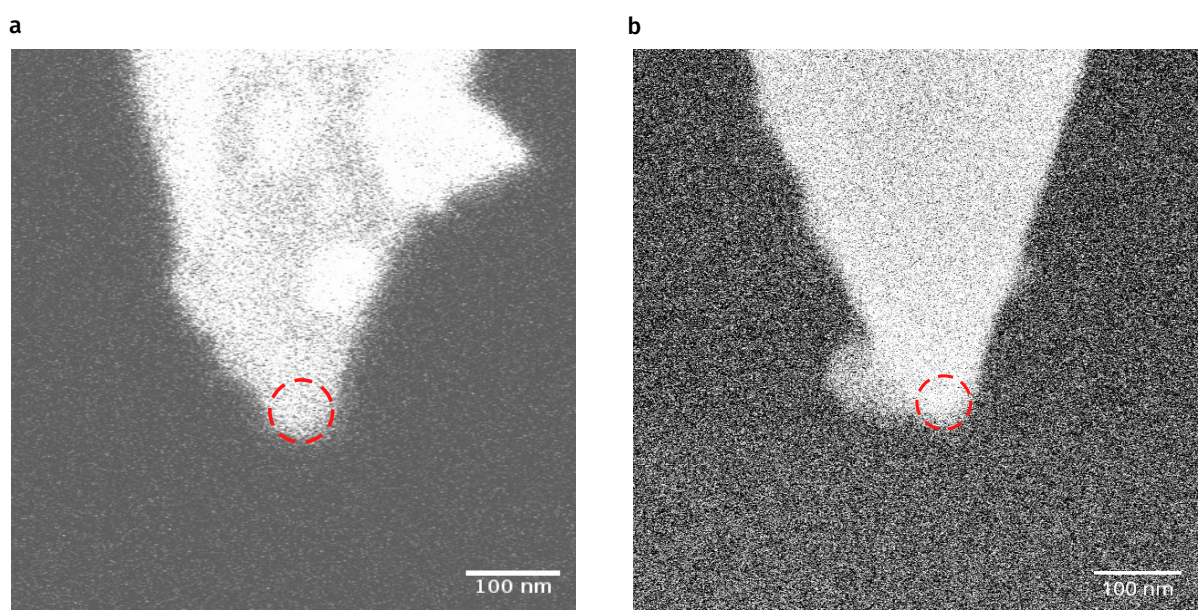

Figure S3. **Magnified SEM images of the AFM probes considered in this study.** The hemispherical shape of the tip is highlighted with a red dashed line and the radius $R$ was estimated to be approximately 33 nm for the gold-coated tip **a** with a calibrated deflection sensitivity of 69.679 $\frac{nm}{V}$ and a spring constant of 0.16136 $\frac{N}{m}$ and 20 nm for the uncoated tip **b** with a calibrated deflection sensitivity of 75.567 $\frac{nm}{V}$ and a spring constant of 0.11670 $\frac{N}{m}$.

## S2.2 AFM background noise subtraction

The raw measurement data was processed by subtracting the background noise, which was determined by fitting the force-distance measurement data with the following function:

$$f_{\text{Bkg}}(d) = p_0 + p_1 d + A\sin(\omega d + \theta) \tag{S6}$$

within the displacement range $d$ between 150 nm to 450 nm, which is far before the tip establishes contact or after it retracts from graphene. The function $f_{\text{Bkg}}$ contains two distinct parts of correction. The linear function $p_0 + p_1 d$, could correct the noise resulting from any alignment issue in the measurement between tip and photodiode. The periodic function $A\sin(\omega d + \theta)$ is to correct the optical interference noise caused by stray light reflected from the supporting material, which is a well-known phenomena taking place during force-displacement measurement of reflective surfaces[S12], [S13]. This noise subtraction method is applied on both freestanding and $SiN_x$ supported graphene for both AFM tips, which yielded consistent results in terms of a stable baseline for further quantification, regardless of the substrate choice and whether the tip approach/retraction measurements. An example of processed force-displacement hysteresis measurement is shown in Fig. S4, where Fig. S4a and S4b correspond to measurements on freestanding and $SiN_x$ supported graphene, respectively. The distance $d = 0$ nm was set to be the point for which the force is minimal, while establishing contact. The noise-subtracted force responses exhibit well-defined zero baseline, with all essential information retained within the contact region, such as repulsive and attractive interaction, as well as the mechanical responses on freestanding and supported graphene upon retraction [S14].

# S3 Additional Discussions

This section contains further discussions that are not directly related to the results in the main text but may help understanding or extending our current work.

## S3.1 Influence of bulk material bandgap

In principle, the vdW repulsion may also be observed on freestanding graphene even if the bulk material is semiconductor or insulator, as long as the inequality (1) in main text holds. Here we theoretically explore the influence of the bandgap of bulk material on the interfacial

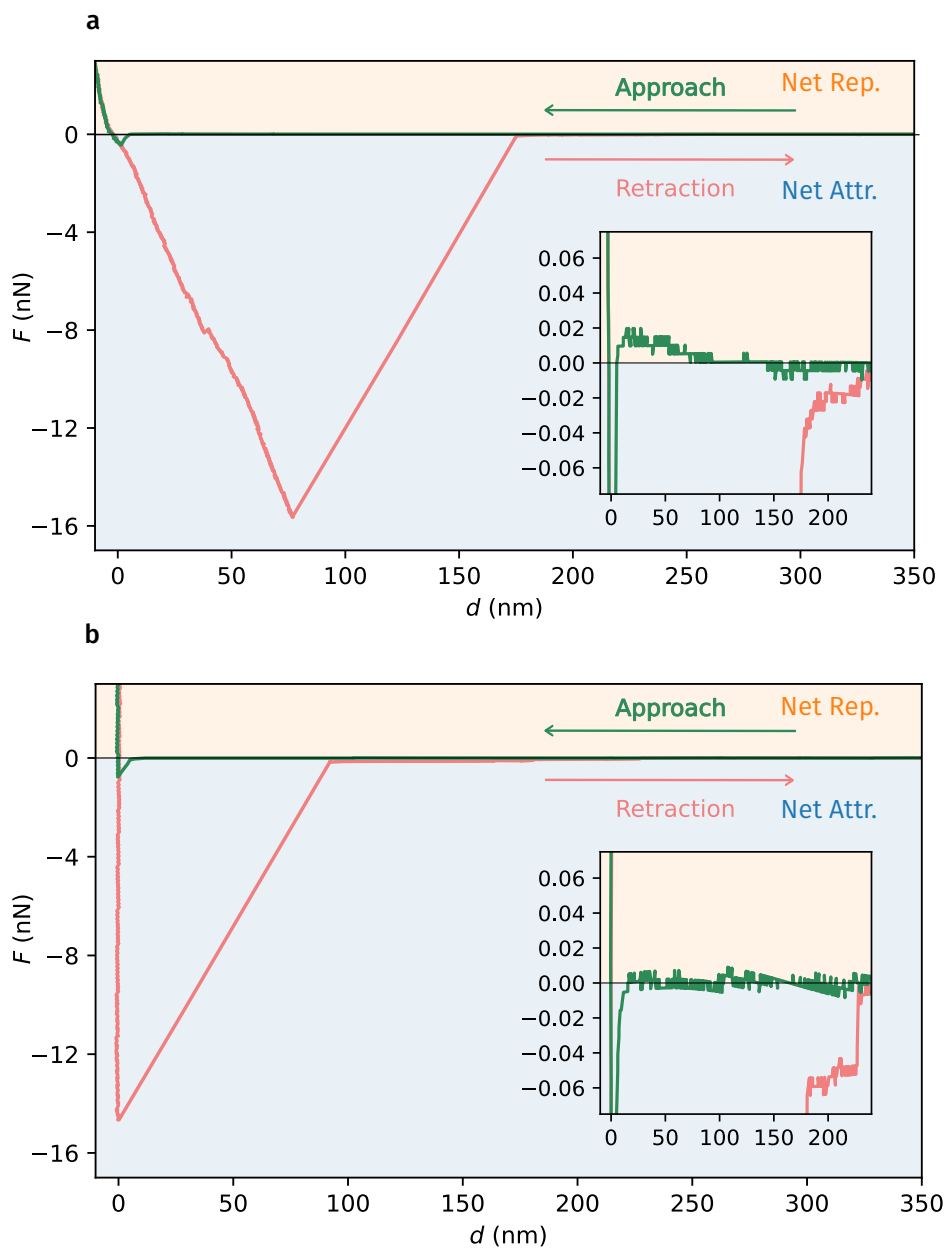

Figure S4. **Representative force-displacement responses after noise substraction of the Au-coated AFM tip.** **a** and **b** correspond to one force-displacement hysteresis measurement on freestanding and $SiN_x$ supported graphene, respectively. The insets magnify the force responses within the contact region. Each measurement contains the approach (green) and retraction (red) responses.

vdW forces in a Vac/Gr/B system, where B is the bulk material with varied bandgap. Fig S5a compared the frequency-dependent dielectric functions for typical metal (Au), semiconductors (GaAs, GaN) and insulator (SiO$_2$) as compared with graphene at $d = 1$ nm. As a general trend, the dielectric response decreases for materials with larger bandgap. As a result, the interaction spectra for Vac/Gr/B systems gradually shift to attractive at lower frequencies, when the bandgap of B increases (Fig. S5b and S5c). The results indicate that on freestanding graphene interface, it may be easiest to observe the vdW repulsion if material B is a metal, compared to semiconductors or insulators.

### S3.2 Influence of 2D material layer number

An intuitive way to control the degree of interfacial forces on 2D materials is to control their layer numbers, which can also be captured by our theoretical framework. We model the many-body vdW interactions by extending the analysis with the layered planar system description from Parsegian's work [S1]. For a system of Vac/NL-Gr/Au where NL-Gr represents $N$-layer graphene stacks, for small $N$, we assume that the polarizability of NL-Gr $\alpha_{\text{NL}}^p$ linearly scales with $N$ [S15], such that

$$\alpha_{\text{NL}}^p = N\alpha_{\text{2D}}^p \tag{S7}$$

where $p$ is either in- or out-of-plane components and $\alpha_{\text{2D}}$ is the two-dimensional electronic polarizability of monolayer graphene.

Using this model, we calculated the distance-dependent total vdW interaction energy $\Phi_{\text{tot}}$ for the Vac/NL-Gr/Au system with varied finite layer numbers, as shown in Fig. S6. As expected, with increasing layer number of graphene, the repulsive interaction decreases. We propose that such change can be probed by AFM on freestanding graphene with controlled layer numbers, which can be prepared using mechanical exfoliation techniques.

### S3.3 Choice of 2D material

Another degree of freedom for controlling the interfacial forces is the choice of 2D material; for example, replacing graphene with MoS$_2$ or hBN. Eq. (S4) shows that $\Phi_{\text{AmB}}^{\text{vdW}}$ depends on:

1. **Dielectric mismatch**: $\Delta_{\text{Am}}\Delta_{\text{Bm}}$

2. **Dielectric anisotropy of $m$**: $g_{\text{m}}$

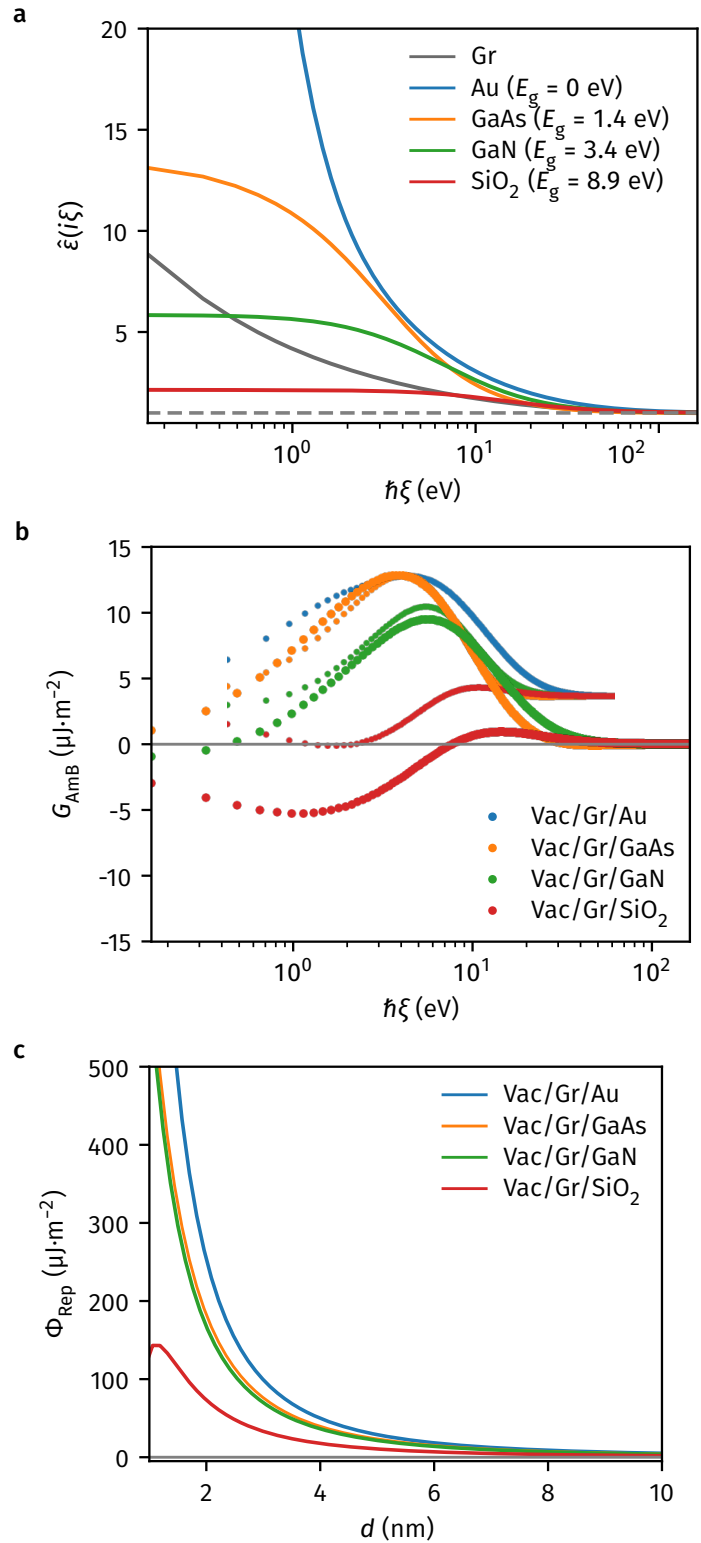

Figure S5. **Influence of bulk material bandgap on the interfacial vdW repulsion a.** Dielectric functions for Au, GaAs, GaN and SiO$_2$ compared with graphene at $d = 1$ nm. **b.** Interaction spectra of Vac/Gr/B systems when B=Au, GaAs, GaN and SiO2, respectively. **c.** Distance-dependent vdW repulsion energy of corresponding system in **b**.

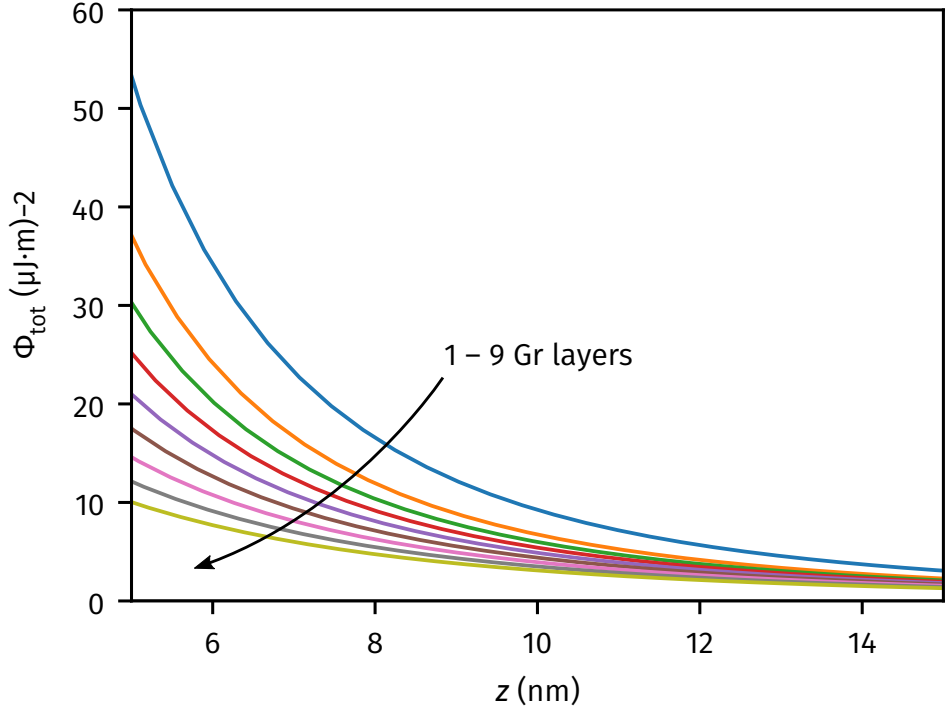

Figure S6. **Total vdW interaction energy for Vac/NL-Gr/Au system with various graphene layer numbers.** With increasing graphene layer thickness, the repulsive response diminishes.

For isotropic medium $m$, $g_m = 1$. The maximal vdW repulsion at a given frequency $\xi$ occurs:

$$\frac{\partial \Delta_{Am}(\xi)\Delta_{Bm}(\xi)}{\partial \hat{\varepsilon}_m(\xi)} = 0 \tag{S8}$$

by definition of $\Delta_{Am}$ and $\Delta_{Bm}$ this is equivalent to $\hat{\varepsilon}_m(\xi) = \sqrt{\varepsilon_A(\xi)\varepsilon_B(\xi)}$. However, using a 2D material as medium, we have $g_m < 1$, and $g_m$ becomes smaller when the bandgap of 2D material decreases[S15]. Therefore the situation is more complex than homogeneous medium. Out theoretical analysis in Fig. S7 shows that, when the effect of 2D anisotropy is considered, hBN actually has stronger repulsive interaction than graphene in the Vac/m/Au systems, due to less screening at lower frequencies. The results indicate that similar interaction behavior may also be observed on other freestanding 2D materials, which we will investigate in future studies.

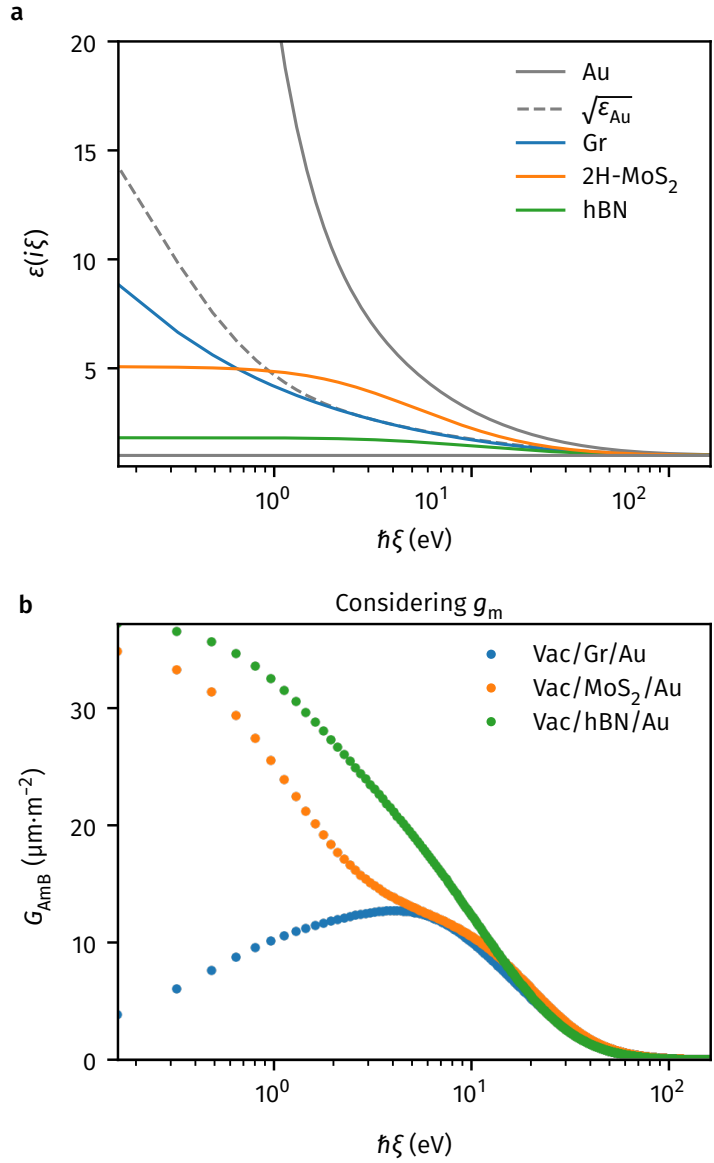

Figure S7. **Influence of choice of 2D material. a**. Frequency-dependent dielectric responses for graphene, MoS$_2$, hBN at $d = 1$ nm. **b** Calculated interaction spectra taking itnto account $g_\mathrm{m}$, revealing that the Vac/hBN/Au system could yield stronger vdW repulsion.


\end{document}